\documentclass[]{spie}

\usepackage{amsmath,amsfonts,amssymb}
\usepackage{graphicx}
\usepackage[colorlinks=true, allcolors=blue]{hyperref}

\title{Which transition can be used in sodium mirrorless lasing for mesospheric magnetometry?}

\author[a,b,*]{Yucheng Yang}
\author[c]{Chunyang Lei}
\author[d]{Kai Guo}
\author[a]{Chi Peng}
\affil[a]{State Key Laboratory of Spatial Datum, Beijing 100020, China}
\affil[b]{Beijing Key Laboratory of Quantum Sensing and Precision Measurement, and Center for Quantum Information Technology, and Institute of Quantum Electronics, Peking University, Beijing 100871, China}
\affil[c]{AthenaEyes Co., Ltd., Changsha, China}
\affil[d]{Institute of Systems Engineering, AMS, Beijing 100141, China}

\authorinfo{Correspondence: yangyucheng@pku.edu.cn}

\begin{document}
\maketitle

%
%
\begin{abstract}
Directed mirrorless lasing from the mesospheric sodium layer has been proposed
as a way to overcome the isotropy of laser-guide-star fluorescence, with
demonstrated cell-scale analogues and a demonstrated stand-off magnetometry
application. Several transition paths on the Na ladder compete for the same
pump photons, and which of them can sustain a population inversion---and under
what pumping format---is usually settled by numerical scans of a multi-level
rate model. Here the same question is answered in closed form, and three
design-relevant results follow from the atomic data alone. (1) A cascade
inversion on $u\to l$ requires $\tau_u/\tau_l>(g_u/g_l)b_{ul}$. The inequality
reproduces the full continuous-wave classification---admitting
$4P_{3/2}\to4S_{1/2}$ (2.21~\textmu m), $4P_{3/2}\to3D_{5/2}$ (9.1~\textmu m)
and the fine-structure companion $4S_{1/2}\to3P_{1/2}$ (1138~nm), excluding
$4D_{5/2}\to4P_{3/2}$ (2.34~\textmu m) and $3D_{5/2}\to3P_{3/2}$
(819~nm)---and shows the 2.34~\textmu m line to miss by only 9\%, which is why
it is transiently accessible. (2) Velocity selectivity reverses the naive
scheme ranking because a Doppler-averaged treatment, exact for one-step
pumping, underestimates two-step pumping by a participation factor of order the
Doppler-to-natural width ratio, $\sim$$10^{2}$; equal division of power between
two pump beams is exactly optimal. (3) The transient window on 2.34~\textmu m
closes after $\sim$2$\tau(4P)\approx210$~ns, set by the reservoir lifetime
rather than by the pump. Each conclusion is traceable to a lifetime, a
branching ratio or a linewidth, and transfers to another species without
rerunning a model.
\end{abstract}


\section{INTRODUCTION}
\label{sec:intro}

Amplified spontaneous emission offers a way around a basic inefficiency of
sodium laser guide stars: the fluorescence they excite in the mesospheric layer
is radiated into the full sphere, and the receiving aperture intercepts almost
none of it~\cite{Holzlohner2010,PedrerosBustos2018}. Were some downward line of
the Na ladder held in population inversion, the illuminated column would itself
become a gain medium and the emission would acquire a preferred direction;
vapor cells supply the proof of principle~\cite{Akulshin2018,Zhang2021}, and
Ref.~\citenum{Akulshin2025} reviews the field.

A companion survey~\cite{YangLei2026map} maps this question numerically: a
ten-level rate model built on NIST atomic data~\cite{NISTASD,Sansonetti2008},
scanned scheme by scheme, returns the set of invertible lines, their
column-gain exponents, the scheme ranking, and the transient behavior of the
2339~nm line. A numerical map, however, states \emph{which} without stating
\emph{why}, and at another species or temperature it must be rerun from
scratch. The present paper treats the same question analytically and
claims none of the survey's results as its own. Its contributions are three
closed forms: an inversion criterion containing nothing but lifetimes,
degeneracies and one branching ratio (Sec.~\ref{sec:criterion}); the validity
boundary of Doppler-averaged rate equations, which turns out to govern the
choice of pump scheme (Sec.~\ref{sec:doppler}); and two format rules fixed by
the atomic constants alone (Sec.~\ref{sec:rules}). Wherever a survey-model
number appears below, it serves as a check on a formula, not as a finding.

\section{AN IRRADIANCE-INDEPENDENT CRITERION FOR CASCADE INVERSION}
\label{sec:criterion}

Let an upper level $u$ feed a lower level $l$ radiatively, let $l$ have no
other significant source, and write $F$ for the total rate at which $u$ is
populated, by whatever mixture of pumping and cascade. The stationary
populations are $n_u=F\tau_u$ and $n_l=F\,b_{ul}\,\tau_l$, with $\tau$ the
radiative lifetimes and $b_{ul}=A_{ul}\tau_u$ the branching ratio of the
connecting line; a positive degeneracy-weighted difference,
$n_u>(g_u/g_l)\,n_l$, then requires
\begin{equation}
\boxed{\;\frac{\tau_u}{\tau_l}\;>\;\frac{g_u}{g_l}\,b_{ul}\;}
\label{eq:criterion}
\end{equation}
with $F$ cancelled. Nothing about the laser survives in
Eq.~(\ref{eq:criterion}): three atomic constants decide, at every irradiance
and under every scheme. (Extra feeding of $l$ replaces $b_{ul}$ by
$F_l/F_u>b_{ul}$ and only hardens the inequality---necessary in general, exact
for a clean cascade.)

\begin{figure}[!htb]
\centering
\includegraphics[width=0.46\textwidth]{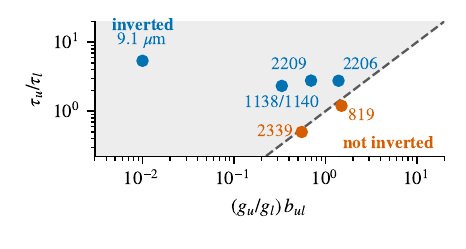}
\caption{The criterion of Eq.~(\ref{eq:criterion}) applied to the candidate
lines. Points above the diagonal (shaded) hold a stationary inversion under any
CW pump; points below cannot. 2339~nm misses by 9\%, 819~nm by 20\%.}
\label{fig:criterion}
\end{figure}

Figure~\ref{fig:criterion} places the candidate lines of the ladder on the
$\bigl((g_u/g_l)b_{ul},\,\tau_u/\tau_l\bigr)$ plane; the lifetimes involved run
16, 19, 38, 52 and 105~ns for $3P$, $3D$, $4S$, $4D$ and $4P$. The two lines
out of $4P_{3/2}$ pass comfortably---2206~nm at $2.76$ versus $1.39$,
9.1~\textmu m at $5.38$ versus $0.010$---since $4P$ outlasts everything beneath
it while feeding the 9.1~\textmu m lower level through a mere 1.7\% branch. The
line $4S_{1/2}\to3P_{1/2}$ at 1138~nm passes at $2.33$ versus $0.333$ even
though no pump touches it; cesium exhibits the analogous companion-line
effect~\cite{Antypas2019}. Two candidates fail: $3D_{5/2}\to3P_{3/2}$ (819~nm)
posts $1.20$ against $1.50$, a 20\% shortfall---before noting that the one
scheme which pumps $3D_{5/2}$ also drives its lower level directly, widening
the gap---and $4D_{5/2}\to4P_{3/2}$ (2339~nm) posts $0.50$ against $0.550$,
because $4P_{3/2}$ both outlasts $4D_{5/2}$ and intercepts 37\% of its decay.

The margins carry information that a pass/fail table cannot. The 2339~nm
shortfall is 9\%---not a factor, not an order of magnitude---and a stationary
condition missed that narrowly announces a transient opportunity: until
equilibrium is established, the line is inverted (Sec.~\ref{sec:rules}). The
survey model can only report ``negative at every irradiance''; the margin
behind that verdict is what Eq.~(\ref{eq:criterion}) exposes.

\section{THE VALIDITY BOUNDARY OF DOPPLER AVERAGING}
\label{sec:doppler}

\begin{figure}[!htb]
\centering
\includegraphics[width=\textwidth]{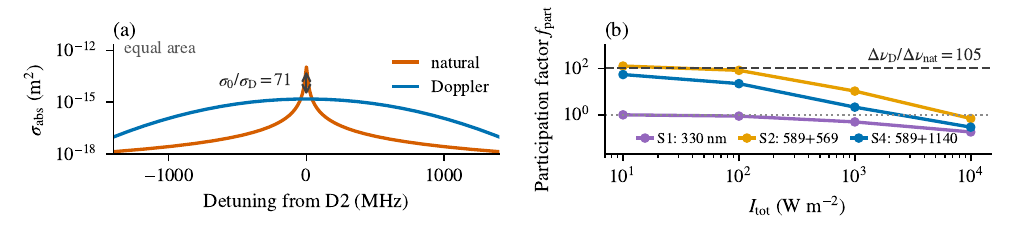}
\caption{(a) The two line-shape choices for the D2 pump at $T=185$~K: equal
areas, peaks apart by the factor of Eq.~(\ref{eq:ratio}). (b) The participation
factor: unity for one step (S1) as Eq.~(\ref{eq:onestep}) demands, near the D2
width ratio (dashed) for two steps (S2, S4), the advantage gone by
$10^{4}$~W\,m$^{-2}$. Dotted: $f_{\rm part}=1$.}
\label{fig:mech}
\end{figure}

Every rate model of mesospheric pumping must pick a line shape for the
absorption cross section
$\sigma_{\rm abs}(\nu)=(g_u/g_l)(\lambda^{2}/8\pi)A_{ul}\,S(\nu)$, with $S$
normalized to unit area~\cite{Hilborn1982}. Two choices of $S$ define the two
models: the Doppler Gaussian $\phi_{\rm D}$ (full width $\Delta\nu_{\rm D}$),
whose peak $\sigma_{\rm D}$ is applied to all atoms---\emph{Doppler
averaging}---or the natural Lorentzian $L$ (full width
$\Delta\nu_{\rm nat}=\Gamma_u/2\pi$), giving each velocity class
$\sigma_0=(\lambda^{2}/2\pi)(g_u/g_l)(A_{ul}/\Gamma_u)$ and summing classes
over the Maxwell distribution. In the mesosphere a class keeps its identity for
the $\sim$35~\textmu s between velocity-changing
collisions~\cite{Holzlohner2010}---hundreds of radiative lifetimes---so the
\emph{class-resolved} model is the physical one there. The two peaks stand in
the ratio
\begin{equation}
\frac{\sigma_0}{\sigma_{\rm D}}=\frac{L(0)}{\phi_{\rm D}(0)}
=\kappa\,\frac{\Delta\nu_{\rm D}}{\Delta\nu_{\rm nat}},
\qquad
\kappa=\frac{1}{\pi\sqrt{\ln 2/\pi}}=0.678,
\label{eq:ratio}
\end{equation}
a factor 71 for D2 at 185~K (Fig.~\ref{fig:mech}(a)), yet both shapes integrate
to the same oscillator strength. Everything below follows from that equal-area
statement.

\textbf{One pump step: no difference, exactly.} In the linear regime the
excited fraction of a class detuned by $\nu$ is
$\sigma_{\rm abs}(\nu)I/(h\nu\,\Gamma_u)$, and summing it against the Maxwell
weight---with the broad Gaussian taken at line center, since
$\Delta\nu_{\rm nat}\ll\Delta\nu_{\rm D}$---gives
\begin{equation}
n_u^{\rm VS}
=\frac{I}{h\nu\,\Gamma_u}\!\int\!\phi_{\rm D}(\nu)\,\sigma_0 L(\nu)\,d\nu
\simeq\frac{I}{h\nu\,\Gamma_u}\,\phi_{\rm D}(0)\,\frac{g_u}{g_l}\frac{\lambda^{2}}{8\pi}A_{ul}
=\frac{\sigma_{\rm D}\,I}{h\nu\,\Gamma_u}=n_u^{\rm DA}:
\label{eq:onestep}
\end{equation}
the participation factor $f_{\rm part}\equiv n_u^{\rm VS}/n_u^{\rm DA}$ equals
one identically, for any line, because the narrow-line sum reassembles the very
area the averaged model spends. The survey model returns $f_{\rm part}=0.983$
for 330~nm pumping at 10~W\,m$^{-2}$; the deficit is the first-order saturation
correction, $I/I_{\rm sat}=10/574=0.017$ against 0.017 observed, with
$I_{\rm sat}=h\nu\,\Gamma_u/\sigma_0$ the per-class saturation irradiance.

\textbf{Two pump steps: a factor
$\sim\Delta\nu_{\rm D}/\Delta\nu_{\rm nat}$.} The identity breaks as soon as
the excitation is a product of two rates. Averaging charges the Doppler price
twice, $n_u^{\rm DA}\propto\sigma_{{\rm D},1}I_1\,\sigma_{{\rm D},2}I_2$;
class resolution confines the excitation to the overlap
$\Delta\nu_{\rm ov}=\Delta\nu_1\Delta\nu_2/(\Delta\nu_1+\Delta\nu_2)$ of the
two Lorentzians but pays the natural-width price twice there, so
\begin{equation}
f_{\rm part}\simeq
\underbrace{\tfrac{\pi}{2}\,\phi_{\rm D}(0)\,\Delta\nu_{\rm ov}}_{\text{addressed fraction}}
\times
\underbrace{\kappa^{2}D_1D_2}_{\text{two enhancements}}
\;\sim\;\kappa^{2}\,\frac{\Delta\nu_{\rm D}}{\Delta\nu_{\rm nat}},
\qquad D_i\equiv\frac{\Delta\nu_{{\rm D},i}}{\Delta\nu_{{\rm nat},i}},
\label{eq:fpart}
\end{equation}
one net power of the width ratio surviving. On the D2 first step the estimate
is $\kappa^{2}\times105=48$; the survey model returns 127 ($589+569$~nm) and 54
($589+1140$~nm) at 10~W\,m$^{-2}$ (Fig.~\ref{fig:mech}(b)), bracketing it. This
factor is what lies behind the survey's reversed scheme ranking near
40~W\,m$^{-2}$~\cite{YangLei2026map}: averaging handicaps exactly the two-step
schemes, by two orders of magnitude, inside the 34--79~W\,m$^{-2}$ that a
20~W-class launch delivers~\cite{Holzlohner2010,PedrerosBustos2020}.

\textbf{The boundary.} All curves of Fig.~\ref{fig:mech}(b) decay because the
addressed class saturates while the averaged model keeps absorbing linearly;
past saturation, extra irradiance buys only a power-broadened width
$\propto\sqrt{1+I/I_{\rm sat}}$. The one-step curve decays as
$(1+I/I_{\rm sat})^{-0.59}$, and the two-step curves drop through unity near
$10^{4}$~W\,m$^{-2}$. Doppler averaging is thus a safe shortcut only above
$\sim$$10^{3}$~W\,m$^{-2}$---and the entire CW-reachable window sits below it.

\section{DESIGN RULES FIXED BY THE ATOMIC CONSTANTS}
\label{sec:rules}

\textbf{Dividing the pump power.} A two-step scheme splits a fixed total
irradiance, $I_1=(1-f)I_{\rm tot}$ and $I_2=fI_{\rm tot}$. Linearity makes the
excitation proportional to $I_1I_2=f(1-f)I_{\rm tot}^{2}$, so
\begin{equation}
f_{\rm opt}=\tfrac{1}{2},
\qquad
\frac{n_u(f)}{n_u(1/2)}=4f(1-f):
\label{eq:split}
\end{equation}
equal division is optimal by construction---independent of scheme, wavelength
and power---and the normalized scan is a parameter-free parabola. Survey scans
confirm both statements to $0.5$\% at 100~W\,m$^{-2}$ (optima $f=0.501$ and
$0.498$ for the two schemes) and to 1.7--4.5\% at $10^{3}$~W\,m$^{-2}$; the
parabola breaks only at $10^{4}$~W\,m$^{-2}$, where the optima drift to 0.56
and 0.40 in opposite directions, the sign set by which step saturates first.
The empirical optimum range the survey quotes is thereby sharpened to an exact
statement with a known expiry irradiance.

\textbf{The usable pulse length.} The 9\% shortfall of
Sec.~\ref{sec:criterion} turns into an asset once the pump is pulsed. When
bichromatic pumping starts abruptly, the upper level of the 2339~nm line fills
within nanoseconds, while its lower level---the $4P_{3/2}$ reservoir whose
excess feeding forbids the stationary inversion---charges up as
$1-e^{-t/\tau_{4P}}$. Setting the reservoir's approach to equilibrium against
the shortfall margin closes the transient window at
$t_{\rm close}\approx2\,\tau(4P_{3/2})=209$~ns, with no pump parameter anywhere
in it; the survey's time integrations close it at 190--290~ns across three
decades of peak irradiance~\cite{YangLei2026map}, the spread being the weak
residual pump dependence the estimate drops. The format rules follow at once:
pulses no longer than $\sim$200~ns, switched on fast relative to the charging
of the intermediate resonance level. How much inversion a pulse wins is set by
its peak irradiance; whether any survives is set by its length against one
atomic lifetime.

\section{CONCLUSIONS}
\label{sec:conclusions}

The division of labor with the companion survey~\cite{YangLei2026map} is
strict. The survey reports what a scanned ten-level model shows: gain exponents
per scheme, the classification table, the vapor-cell anchor, transient curves.
This paper reports why those outcomes could not have been otherwise---the
criterion $\tau_u/\tau_l>(g_u/g_l)b_{ul}$ with its 9\% and 20\% exclusion
margins, the equal-area argument fixing where Doppler averaging fails and by
how much, and the format rules $f_{\rm opt}=1/2$ and
$t_{\rm close}\approx2\tau(4P)$. No survey result is re-derived here as a
claim; its numbers enter only as checks, and they pass. The closed forms are
also the transferable part: evaluating them for potassium or iron takes
spectroscopic tables, not a model.

\acknowledgments
This work was supported by the National Natural Science Foundation of China
under Grant No.~62301377.

\bibliography{refs}

@article{Holzlohner2010,
  author  = {Holzl{\"o}hner, R. and Rochester, S. M. and Bonaccini Calia, D. and Budker, D. and Higbie, J. M. and Hackenberg, W.},
  title   = {Optimization of cw sodium laser guide star efficiency},
  journal = {Astronomy \& Astrophysics},
  volume  = {510},
  pages   = {A20},
  year    = {2010}
}

@article{PedrerosBustos2018,
  author  = {Pedreros Bustos, F. and Bonaccini Calia, D. and Budker, D. and Centrone, M. and Hellemeier, J. and Hickson, P. and Holzl{\"o}hner, R. and Rochester, S.},
  title   = {Remote sensing of geomagnetic fields and atomic collisions in the mesosphere},
  journal = {Nature Communications},
  volume  = {9},
  pages   = {3981},
  year    = {2018}
}

@article{Akulshin2025,
  author  = {Akulshin, A. M. and Budker, D. and Pedreros Bustos, F. and Dang, T. and Klinger, E. and Rochester, S. M. and Wickenbrock, A. and Zhang, R.},
  title   = {Remote detection optical magnetometry},
  journal = {Physics Reports},
  volume  = {1106},
  pages   = {1--32},
  year    = {2025}
}

@article{Akulshin2018,
  author  = {Akulshin, A. M. and Pedreros Bustos, F. and Budker, D.},
  title   = {Continuous-wave mirrorless lasing at 2.21~{\textmu}m in sodium vapors},
  journal = {Optics Letters},
  volume  = {43},
  number  = {21},
  pages   = {5279--5282},
  year    = {2018}
}

@article{Antypas2019,
  author  = {Antypas, D. and Tretiak, O. and Budker, D. and Akulshin, A.},
  title   = {Polychromatic, continuous-wave mirrorless lasing from monochromatic pumping of cesium vapor},
  journal = {Optics Letters},
  volume  = {44},
  number  = {14},
  pages   = {3657--3660},
  year    = {2019}
}

@article{Zhang2021,
  author  = {Zhang, R. and Klinger, E. and Pedreros Bustos, F. and Akulshin, A. and Guo, H. and Wickenbrock, A. and Budker, D.},
  title   = {Stand-off magnetometry with directional emission from sodium vapors},
  journal = {Physical Review Letters},
  volume  = {127},
  pages   = {173605},
  year    = {2021}
}

@article{Sansonetti2008,
  author  = {Sansonetti, J. E.},
  title   = {Wavelengths, transition probabilities, and energy levels for the spectra of sodium ({Na~I}--{Na~XI})},
  journal = {Journal of Physical and Chemical Reference Data},
  volume  = {37},
  number  = {4},
  pages   = {1659},
  year    = {2008}
}

@misc{NISTASD,
  author       = {{NIST ASD Team}},
  title        = {{NIST} Atomic Spectra Database},
  howpublished = {\url{https://physics.nist.gov/asd}},
  note         = {National Institute of Standards and Technology, Gaithersburg, MD; retrieved 16 July 2026}
}

@article{Hilborn1982,
  author  = {Hilborn, R. C.},
  title   = {Einstein coefficients, cross sections, $f$ values, dipole moments, and all that},
  journal = {American Journal of Physics},
  volume  = {50},
  pages   = {982},
  year    = {1982}
}

@article{PedrerosBustos2020,
  author  = {Pedreros Bustos, F. and Holzl{\"o}hner, R. and Rochester, S. and Bonaccini Calia, D. and Hellemeier, J. and Budker, D.},
  title   = {Frequency chirped continuous-wave sodium laser guide stars: modeling and optimization},
  journal = {Journal of the Optical Society of America B},
  volume  = {37},
  pages   = {1208},
  year    = {2020}
}

@misc{YangLei2026map,
      title={Population-inversion map of the mesospheric sodium ladder: continuous-wave and pulsed pumping schemes for directed emission}, 
      author={Yucheng Yang and Chunyang Lei and Kai Guo and Chi Peng and Zongpeng Pan},
      eprint={2608.09037},
      archivePrefix={arXiv},
      primaryClass={physics.optics},
      howpublished = {\url{https://arxiv.org/abs/2608.09037}},
      year={2026}
}
\bibliographystyle{spiebib}

\end{document}